\documentclass[prr,twocolumn,floatfix,superscriptaddress,footinbib,  longbibliography, numbers, merge, reprint]{revtex4-2}
\usepackage{graphicx}
\usepackage{amsmath}
\usepackage{amssymb}
\usepackage{booktabs}
\usepackage{color}
\usepackage{natbib}

\usepackage[utf8]{inputenc}
\usepackage[tight,sf]{subfigure}%

\def\vec #1{{\bf #1}}

\newcommand{\be}{\begin{eqnarray}}
\newcommand{\ee}{\end{eqnarray}}

\newcommand{\todo}[1]{\textcolor{black}{#1}}

\begin{document}
	
\title{Discharging dynamics of topological batteries}

\author{Vishal P. Patil}
\thanks{V.P.P. and \v{Z}.K. contributed equally to this work.}
\affiliation{Department of Mathematics, Massachusetts Institute of Technology, 77 Massachusetts Avenue, Cambridge,~MA~02139, USA}

\author{\v{Z}iga Kos}
\thanks{V.P.P. and \v{Z}.K. contributed equally to this work.}
\affiliation{Department of Mathematics, Massachusetts Institute of Technology, 77 Massachusetts Avenue, Cambridge,~MA~02139, USA}
\affiliation{Faculty of Mathematics and Physics, University of Ljubljana, Jadranska 19, 1000 Ljubljana, Slovenia}

\author{Miha Ravnik}
\affiliation{Faculty of Mathematics and Physics, University of Ljubljana, Jadranska 19, 1000 Ljubljana, Slovenia}
\affiliation{J. Stefan Institute, Jamova 39, 1000 Ljubljana, Slovenia}

\author{J\"orn Dunkel} 
\affiliation{Department of Mathematics, Massachusetts Institute of Technology, 77 Massachusetts Avenue, Cambridge,~MA~02139, USA}

\date{\today}

\begin{abstract}
Topological constraints have long been known to provide efficient mechanisms for localizing and storing energy across a range of length scales, from knots in DNA  to turbulent plasmas. Despite recent theoretical and experimental progress on the preparation of topological states, the role of topology in the discharging dynamics is not well understood. Here, we investigate robust topological energy release protocols in two archetypal soft systems through simulations of 238 knotted elastic fibers and 3D liquid crystals across a range of different topologies. By breaking the elastic fiber or switching the liquid crystal surface anchoring, such topological batteries can perform mechanical work or drive fluid flows. Our study reveals  topologically resonant states for which energy release becomes superslow or superfast. Owing to their intrinsic stability we expect such tunable topological batteries to have broad applications to storage and directed release of energy in soft matter.
\end{abstract}

\maketitle


Topological protection provides a robust means for storing and controlling energy, an effect  widely used in a variety of biological and physical systems~\cite{virnau2006intricate,moffatt1969,senyuk2013topological, goldstein2012topological, ferraro2018high, ren2010berry}. On small scales, knotted topologies play important functional roles~\cite{kauffman2001knots} in the behavior of tangled DNA, proteins and polymers~\cite{marenduzzo2009dna,2006Grosberg,rawdon2015subknots,stasiak1996electrophoretic,gonzalez1999global}. In continuum systems, foundational work on topology has revealed the origin of helicity conservation in classical~\cite{moffatt1969,kleckner2013} and complex fluids~\cite{kedia2013tying}, and the dynamo effect in turbulent plasmas~\cite{taylor1974relaxation, ricca2017groundstate}. Structured continua such as liquid crystal fluids are a rich source of emergent topological phenomena, from interacting defects~\cite{giomi2017cross,muvsevivc2006two} to knotted field configurations~\cite{tkalec2011reconfigurable,senyuk2013topological,machon2016global}. Recent experimental advances in mechanical lattices~\cite{sun2012surface, rocklin2017transformable} and soft robotics~\cite{Kim:2018aa}, bring the question of topologically tunable designer materials into the experimentally accessible realm. Although the study of topological modes~\cite{paulose2015topological} has dramatically improved our understanding of soft matter, harnessing topology to perform useful work such as driving flows~\cite{giomi2017cross,sengupta2013,matthews2010complex}  continues to present fundamental challenges.

\par

Here we study how topology affects energy release dynamics in knotted elastic fibers~\cite{audoly2007elastic,jawed2015untangling,patil2020topological} and nematic liquid crystals~\cite{tkalec2011reconfigurable}, demonstrating two distinct realizations of a topological battery. In both cases, topology mediated buckling and instability phenomena underlie the discharging rates and functional capabilities of the batteries. Knotted filaments present an intuitive mechanical realization of a topological battery: By initializing closed elastic loops in tight knotted states~\cite{cantarella2012shapes,katritch1996geometry,jawed2015untangling} with varying twist~\cite{moffatt1990energy,chui1995energy}, energy may be stored robustly  (Fig~\ref{knots1}).  Cutting the knot at the point of maximum stress results in a controlled topology dependent energy discharge. Transferring this idea to liquid crystals, nematic batteries can be realized by imprinting topologically non-trivial field configurations, which are energetically stabilized through the anchoring of the nematic orientational director field to a colloidal surface~\cite{tkalec2011reconfigurable,senyuk2013topological,machon2016global}. Energy is then released by optically changing the anchoring profile. Using 3D numerical simulations, we explore the energy discharge dynamics in elastic fibers and liquid crystals across a range of different knot types and topological charges. For both systems, we find special topologically resonant states that are characterized by superslow or superfast energy release, exemplifying control of discharging dynamics through topology.

\begin{figure*}[t!]
	\includegraphics[width=1.85\columnwidth]{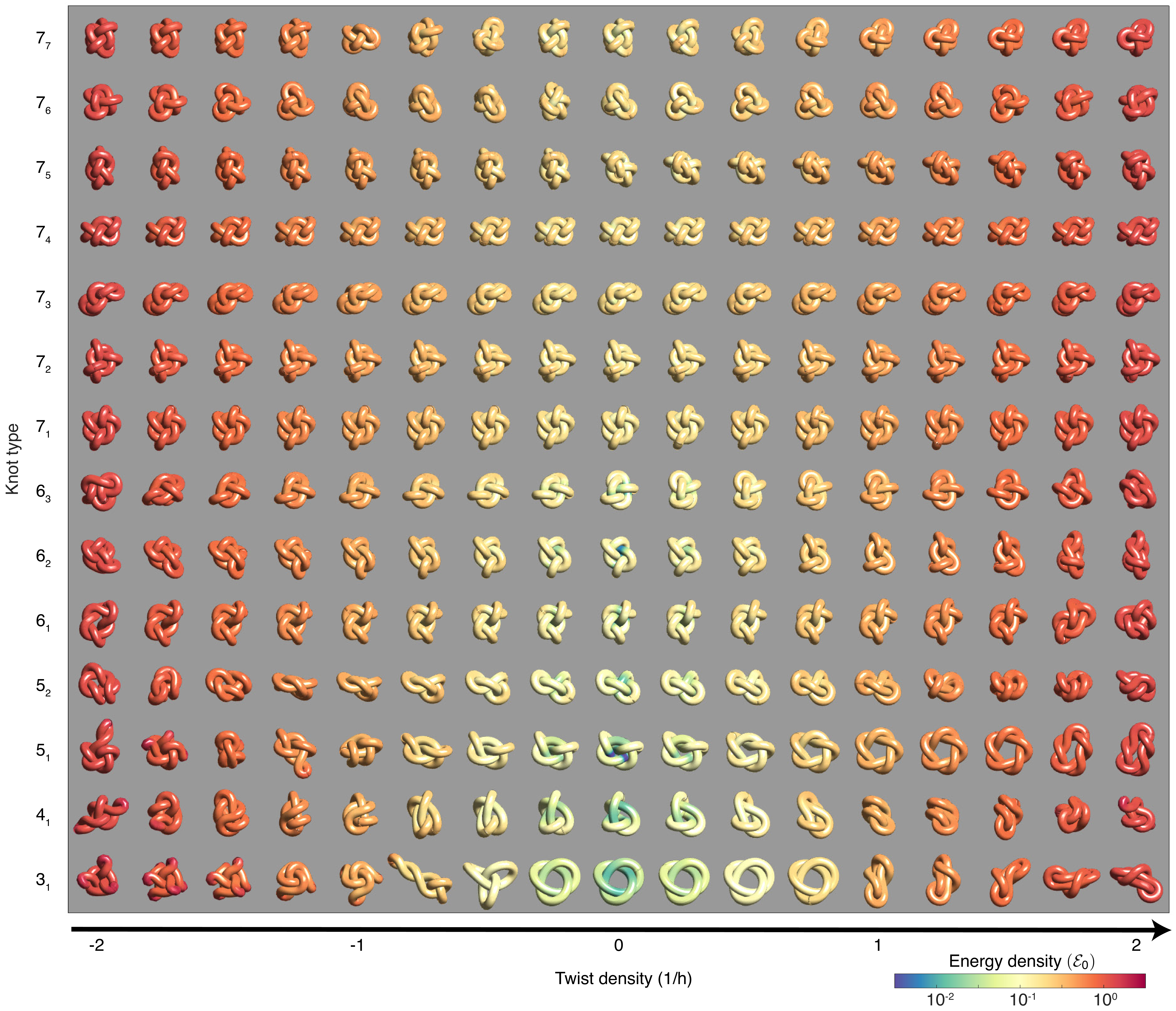}
	\caption{Knotted topological batteries were simulated for 238 initial configurations, representing 17 different twist densities for 14 knot types. Simulation parameters: $\alpha^{-1} = 1.3,\; \gamma = 5,\; L/h = 50$.	\label{si_238}} 
\end{figure*}
\section{Energy release dynamics in elastic knots}

\subsection{Kirchhoff model with contact friction}

Tying a knot in an elastic fiber and fusing together the free ends produces our first example of a topological battery. By twisting the free ends before fusing, knotted batteries can be studied along the two axes of knot type and twist density (Fig~\ref{si_238}). We describe this system using the Kirchhoff model for elastic rods together with contact friction~\cite{patil2020topological,audoly2013discrete}. The fibers have natural length $L$, radius $h$, and circular cross section with moment of inertia $I = \pi h^4 /4$ and cross sectional area $A = \pi h^2$. The state of the fiber is defined by its centerline curve $\boldsymbol{x}(s)$, and an orthonormal material frame, $\{\mathbf{d}_1(s),\mathbf{d}_2(s),\mathbf{d}_3(s)\}$ constrained by $\boldsymbol{x}' \times \mathbf{d}_3 = \mathbf{0}$, where $s$ is the arc length parameter of the unstretched fiber. The elastic energy depends on the geometric curvature $\kappa$, twist density $\theta' = \mathbf{d_1}'\cdot\mathbf{d_2}$, and stretch $|\boldsymbol{x}'|$
\begin{align}\label{kirchE}
\mathcal{E} = \frac{E_bA}{2}\int_0^L ds \left[\frac{1}{4}h^2(\kappa^2 + \alpha \theta'^2) + \gamma(|\boldsymbol{x}'| - 1)^2 \right]
\end{align}
where $E_b$ is the bending modulus, $E=\gamma E_b$ is the Young's modulus and $ \nu=\alpha^{-1} -1$ is the Poission's ratio.  To describe the dynamics, we define $\boldsymbol{\Omega} = \left(\dot{\mathbf{d}}_1\cdot\mathbf{d_2}\right)\mathbf{d}_3$. The governing equations follow from extremizing (\ref{kirchE}) and assuming viscous damping forces and friction dominate inertial terms
\begin{subequations}
\label{Kirch}
\begin{eqnarray}
\mathbf{F'} &=& -\eta A \dot{\boldsymbol{x}}''  -\boldsymbol{f}_\text{fric}, 
\\
\mathbf{M}' + \mathbf{\boldsymbol{x}'\times F} &=& -2\eta I \left(\boldsymbol{\Omega}''\cdot\mathbf{d_3}\right)\mathbf{d_3}
\end{eqnarray}
\end{subequations}
$\mathbf{F}$ and $\mathbf{M}$ are the internal force and moment in the rod respectively, $\boldsymbol{f}_\text{fric}$ is a friction force density, and $\eta$ is a damping parameter. Constitutive laws yield expressions for $\mathbf{F}$ and $\mathbf{M}$
\begin{align*}
\mathbf{M} &=E_bI\left(\mathbf{d_3}\times\mathbf{d_3}'+\alpha\theta'\mathbf{d_3}\right) \\
\mathbf{F}\cdot\mathbf{d_3} &= EA(|\boldsymbol{x}'|-1)
\end{align*}
\todo{The friction force density is based on a friction model that has been experimentally validated~\cite{patil2020topological}
\begin{align*}
\boldsymbol{f}_\text{fric} = \eta_f A \int ds' \, \Theta\left( 2h - |\boldsymbol{x}(s) - \boldsymbol{x}(s')| \right) \frac{\dot{\boldsymbol{x}}(s) - \dot{\boldsymbol{x}}(s') }{ |\boldsymbol{x}(s) - \boldsymbol{x}(s')|^3}
\end{align*}
In contrast to~\cite{patil2020topological}, we take $\eta_f/\eta = 0.1$ (instead of setting $\eta_f = \eta$), and neglect the effect of torsional friction.} The intrinsic energy density, $\mathcal{E}_0 = E_b h^2$, and energy, $E_0 = E_b h^2 L$, of the system correspond to the bending energy associated with $\kappa\sim 1/h$. We choose an intrinsic timescale, $T_0= 2L\eta/hE_b$, as an intermediate scale between the relxation times of pure twist, $T_{tw} = \eta/\alpha E_b$, and pure bending $T_b = 2L^2 \eta / \pi^2 h^2 E_b$. The model applies to a range of different materials;  a  specific  candidate system are thin lubricated silicone fibers with $\alpha^{-1} \sim 1.3, \; \gamma \sim 5, \; \eta \sim 10\,$kPa$\cdot$s, $E_b \sim 10\,$MPa.

\begin{figure*}[t!]
	\includegraphics[width=1.85\columnwidth]{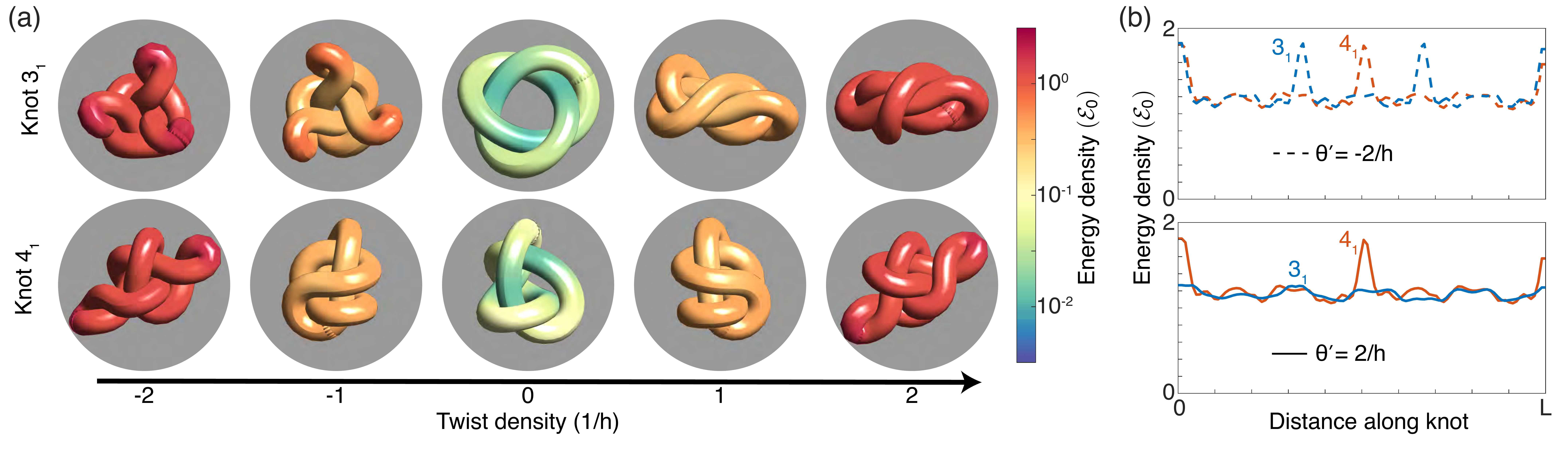}
	\caption{Charging topological batteries with knot tying and twist, illustrated for two knot types.
		(a)~Energy storage in elastic batteries depends on both knot type and initial twist density. The chiral trefoil knot ($3_1$, top) exhibits different buckled states at positive and negative twist whereas the achiral figure-of-eight knot ($4_1$, bottom) does not.  
		(b)~Chirality-dependent buckling leads to different energy density profiles for the trefoil knot when charged with positive twist ($\theta' = 2/h$) and negative twist ($\theta' = -2/h$). Simulation parameters: $\alpha^{-1} = 1.3,\; \gamma = 5,\; L/h = 50$.
	} 
	\label{knots1}
\end{figure*}

\subsection{Charging}

Charging knotted filaments with twist results in a series of topology mediated buckling instabilities (Fig~\ref{knots1}a). The twist in the buckled state is typically lower than the initial twist supplied to the pre-buckled state~\cite{goriely2006twisted}. The evolution of the total twist in the fiber, $Tw$, follows from relating spatial and time derivatives of the frame
\begin{align}\label{twist_transport}
\frac{d}{dt}Tw = \left[\boldsymbol{\Omega}\cdot\mathbf{d}_3\right]^L_0 + \int_0^L ds\; \mathbf{d}_3 \cdot \left(\mathbf{d}_3' \times \dot{\mathbf{d}}_3\right)
\end{align}
When the fiber is closed, the first term on the right hand side vanishes. Although initializing a fiber with a specific twist requires a time dependent torque application protocol, the amount of buckling provides an indicator of the twist charge within the battery, for a given knot type. For example, twist produces initial battery states with varying post-buckled geometries (Fig~\ref{knots1}a) and energy densities (Fig~\ref{knots1}b). Additionally, twist is sensitive to certain underlying topological properties. In our simulations, chiral knots generally exhibit handedness-dependent buckling (Fig~\ref{knots1}a, top), whereas achiral knots tend to buckle independently of twist handedness (Fig~\ref{knots1}a, bottom).

\subsection{Discharging dynamics and topological resonance}

\begin{figure*}[t!]
	\includegraphics[width=1.85\columnwidth]{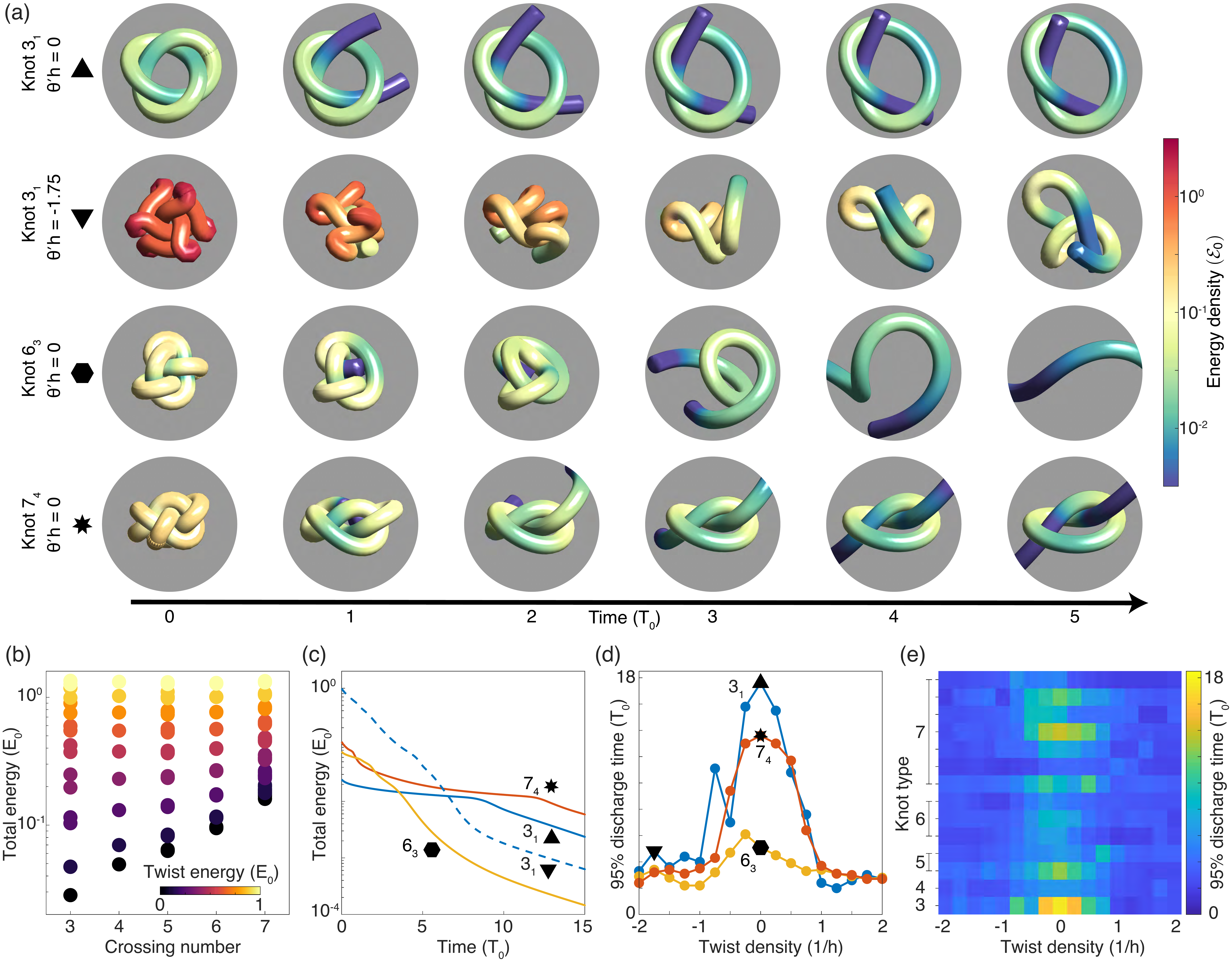}
	\caption{Discharging knotted elastic fibers.
		(a)~The relaxation dynamics depend strongly on knot type and twist state (Movie~S1~\cite{suppmat}). 
		(b)~At low twist, the initial knot energy varies with knot type, while high twist negates the effect of topological changes.
		(c)~Energy is released in phases as intermediate metastable topological states form and untie. Topology dependent obstructions cause certain knot types to untie more slowly at zero twist ($3_1, 7_4$). Higher twist states typically discharge faster but the final discharge rates are set by the relaxed length of the fiber.  
		(d)~Topological resonances occur predominantly at low twist. 
		(e)~High twist leads to fast untying for all knot types. At scales where bending dominates, topology dependent resonance effects become visible. Total number of initial knot configurations simulated for (e) is 238 (Fig~\ref{si_238}) using the algorithm from Refs.~\cite{patil2020topological, cantarella2012shapes}.
	} 
	\label{knots2}
\end{figure*}

As knotted mechanical batteries unravel, the competition between bending and twisting energies reveals a landscape of topological resonances, where certain initial states lead to super slow energy discharge. After being broken at the point of maximum stress, different transient, metastable states occur (Fig~\ref{knots2}a, Movie~S1~\cite{suppmat}). The discharge dynamics separate into a bending dominated regime and a twist dominated regime, as evident from the initial state of the battery (Fig~\ref{knots2}b). The crossover between the two regimes corresponds to the scaling of strains associated with twisting and bending. From (\ref{kirchE}), the characteristic strains for tightly knotted configurations are $\varepsilon_b \sim h\kappa \sim 1$ for bending and $\varepsilon_{tw} \sim h\theta' \sim hTw/L$ for twisting. As a result, twist dominates the discharge dynamics at high twist densities with $\theta' h > 1$ (Fig~\ref{knots2}c-e). Since the effects of varying knot type are more naturally related to the bending energy, twist can be thought of as washing out topology. Indeed, at high twist, $\theta' h>1$, the batteries discharge quickly, independent of knot type (Fig~\ref{knots2}d,e). By contrast, at low twist, knot topology essentially determines the discharging dynamics (Fig~\ref{knots2}e). In particular, select knots exhibit extremely long discharge times (Fig~\ref{knots2}d,e). We can explain these slow topological resonances by considering the mechanisms by which knots release twist and bending energy. Bending forces point in the $\mathbf{d}_3'$ direction, which lies in the fiber's local osculating plane, spanned by $\mathbf{d}_3$ and $\mathbf{d}_3'$. From (\ref{twist_transport}), twist changes when $\dot{\mathbf{d}}_3$ has a component in the $\mathbf{d_3}\times \mathbf{d_3}'$ direction; twist relaxation therefore pushes the fiber out of plane. The topologically resonant slow knots can thus be thought of as maximally non-planar and therefore self-confining.

\begin{figure*}[t!]
	\includegraphics[width=1.85\columnwidth]{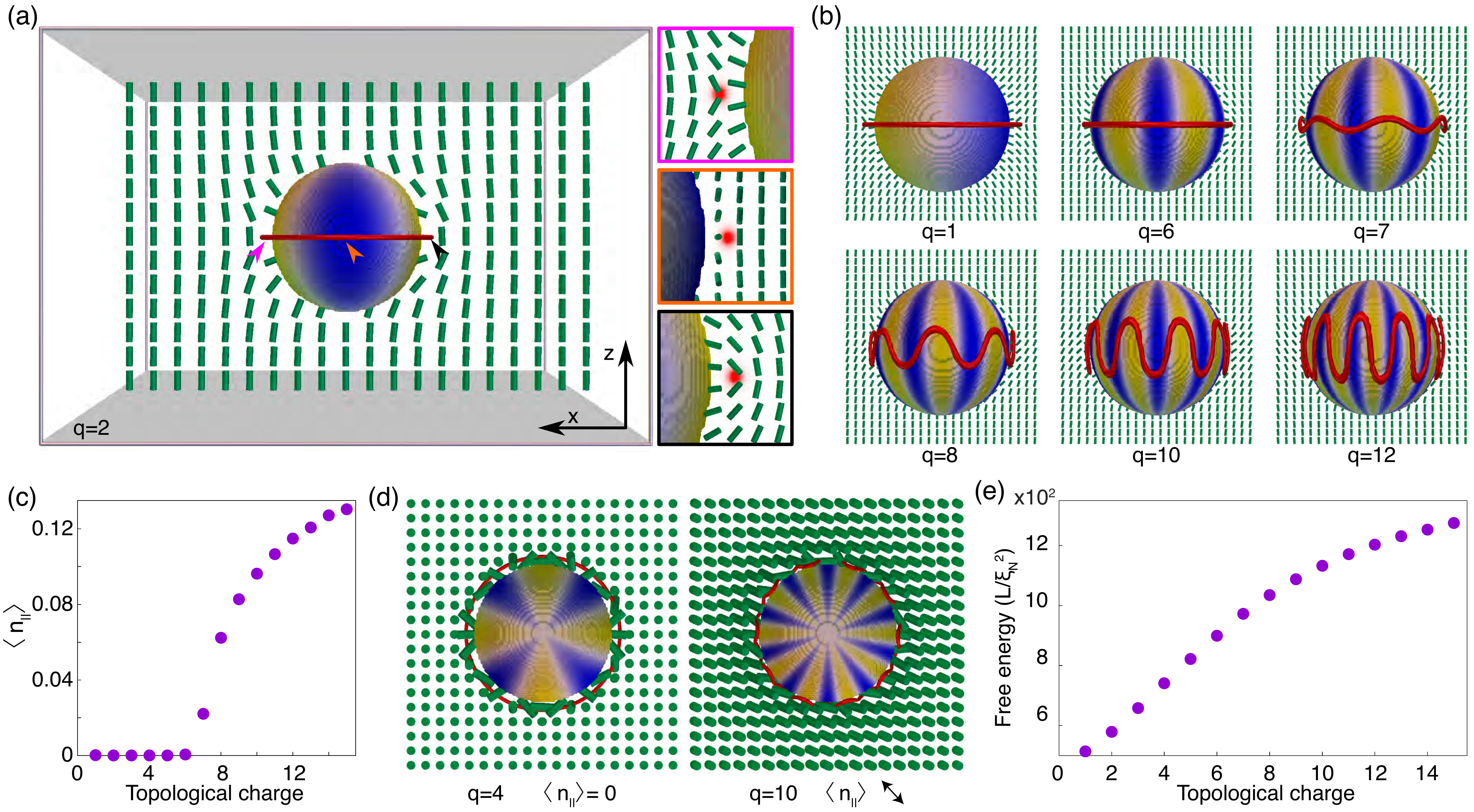}
	\caption{Charging topological batteries in nematic fluids.
		(a)~The battery consists of a spherical colloidal particle confined between two plates (grey), with topologically non-trivial anchoring profile, $\mathbf{n}_q(\theta,\phi)$, resulting in an effective topological charge, $q$, for the particle. The particle is colored from blue to yellow as the horizontal component of $\mathbf{n}_q$ varies from -1 to 1. The director field (green rods) contains a defect ring (red isosurface at $S=0.35$) of varying cross-sectional structure (color coded insets).
		(b)~The defect ring buckles for $q\geq 7$ into a loop with $q$ maxima and minima.
		(c,d)~Above the instability threshold ($q\geq 7$), a non-zero mean director component emerges in the horizontal plane ($xy$) of the sample.
		(e)~The total free energy of the nematic colloid is dependent on the topological charge of the nematic structure. See SI for simulation parameters.
	} 
	\label{nem1}
\end{figure*}

\section{Energy release dynamics in nematic fluids}

\subsection{Nematodynamic model}

Nematic fluids containing spherical colloidal particles~\cite{tkalec2011reconfigurable} enable another construction of a topological battery. The effective topological charge of the particle is determined by its prescribed liquid crystal anchoring profile, which may be optically changed~\cite{ONeillM_JPhysDApplPhys33_2000,SunJ_OptLett38_2013} to release energy. We model the nematic system with a tensor order parameter~$\mathbf{Q}$. The top eigenvalue and eigenvector correspond to the degree of order $S$ and the director $\vec{n}$. The free energy, $F = \int f dV$, is given by
$$
f=\frac{A}{2}Q_{ij}Q_{ji}+\frac{B}{3}Q_{ij}Q_{jk}Q_{ki}+\frac{C}{4}(Q_{ij}Q_{ji})^2 + \frac{L}{2}(\partial_kQ_{ij})^2
$$
Derivatives of $Q_{ij}$ in the free energy density describe the effective elastic behavior of the director field, where $L$ is the elastic constant, and $A$, $B$, and $C$ are parameters that tune the nematic phase behavior. . The model has intrinsic length scale $\xi_\text{N}=\sqrt{L/(A+BS_\text{eq}+\tfrac{9}{2}CS_\text{eq}^2)}$ and time scale $\tau_\text{N}=\xi_\text{N}^2/\Gamma L$, where $\Gamma$ is the rotational viscosity parameter. Our 3D nematodynamic simulations~\cite{kos2017nematodynamics} are based on the Beris-Edwards formulation of the Q-tensor dynamics~\cite{book-Beris}
\begin{align}
&\left(\partial_t+u_k\partial_k\right)Q_{ij}+\Omega_{ik}Q_{kj}-Q_{ik}\Omega_{kj}=\Gamma H_{ij}+\frac{2\chi}{3}D_{ij}\nonumber \\
&+\chi (D_{ik}Q_{kj}+Q_{ik}D_{kj})-2\chi\bigl(Q_{ij}+\frac{\delta_{ij}}{3}\bigr)Q_{kl}W_{kl}\label{eq:BE}
\end{align}
where $\vec{u}$ is fluid velocity, $\chi$ is the alignment parameter, and $\Gamma$ is the rotational viscosity coefficient. The molecular field $H_{ij}$ drives the system towards equilibrium, $H_{ij}=-({\delta F}/{\delta Q_{ij}})_\text{tr}$, where $(.)_\text{tr}$ denotes the traceless part. $D_{ij}$ and $\Omega_{ij}$ are the symmetric and antisymmetric parts of the velocity gradient tensor $W_{ij}=\partial_iu_j$. We model the nematic as an incompressible fluid with stress tensor 
\begin{align}
\sigma_{ij}&=2\chi\left(Q_{ij}+\frac{\delta_{ij}}{3}\right)Q_{kl}H_{kl}-\chi H_{ik}\left(Q_{kj}+\frac{\delta_{kj}}{3}\right) \nonumber \\
&\phantom{={}}-\chi\left(Q_{ik}+\frac{\delta_{ik}}{3}\right)H_{kj}-\partial_iQ_{kl}\frac{\delta F}{\delta\partial_jQ_{kl}}\label{eq:passive}\\
&\phantom{={}}+Q_{ik}H_{kj}-H_{ik}Q_{kj}+2\eta D_{ij}-p\delta_{ij},\nonumber
\end{align}
where $p$ is the fluid pressure and $\eta$ is the isotropic viscosity. Defining the intrinsic length scale $\xi_\text{N}=\sqrt{L/(A+BS_\text{eq}+\tfrac{9}{2}CS_\text{eq}^2)}$ and time scale $\tau_\text{N}=\xi_\text{N}^2/\Gamma L$, the particle radius is set to $R=52.5\xi_\text{N}$ and the director field has relaxation timescale $\tau_\text{d}=R^2/\Gamma L=2760\tau_\text{N}$.
The phase parameters are set to $A=-0.19\,L/\xi_\text{N}^2$, $B=-2.34\,L/\xi_\text{N}^2$ and $C=1.91\,L/\xi_\text{N}^2$, The nematic is in the alignment regime $\chi=1$, with isotropic viscosity $\eta=1.38\,\xi_\text{N}^2/L\tau_\text{N}$. Our 3D simulations were performed on a $200\times 200\times 150$ mesh with grid resolution $\Delta x=1.5\,\xi_\text{N}$ and time resolution $\Delta t=0.057\,\tau_\text{N}$. This Q-tensor based formulation of nematohydrodynamics has been particularly effective for predicting and explaining experimental data measured in studies of entangled defect nematic lines in collidal systems~\cite{tkalec2011reconfigurable,muvsevivc2006two}, nematic colloids with variation of the particle or interface shape~\cite{LuoY_NatCommun9_2018,MartinezA_NatMater13_2014}, patterned nematic interfaces~\cite{WangX_SoftMatter13_2017}, and microfluidics using nematic liquid crystals~\cite{giomi2017cross}.
\todo{
Such mesoscopic approaches can be directly expanded with other contributions and fluid mechanisms, including multiple elastic constants, driving of surface anchoring, and the effects of other possible mechanical or external fields. In addition, surface-contributed time scales could also be controlled through in-plane switching methods~\cite{book-Chigrinov}.}

\begin{figure*}[t!]
	\includegraphics[width=1.85\columnwidth]{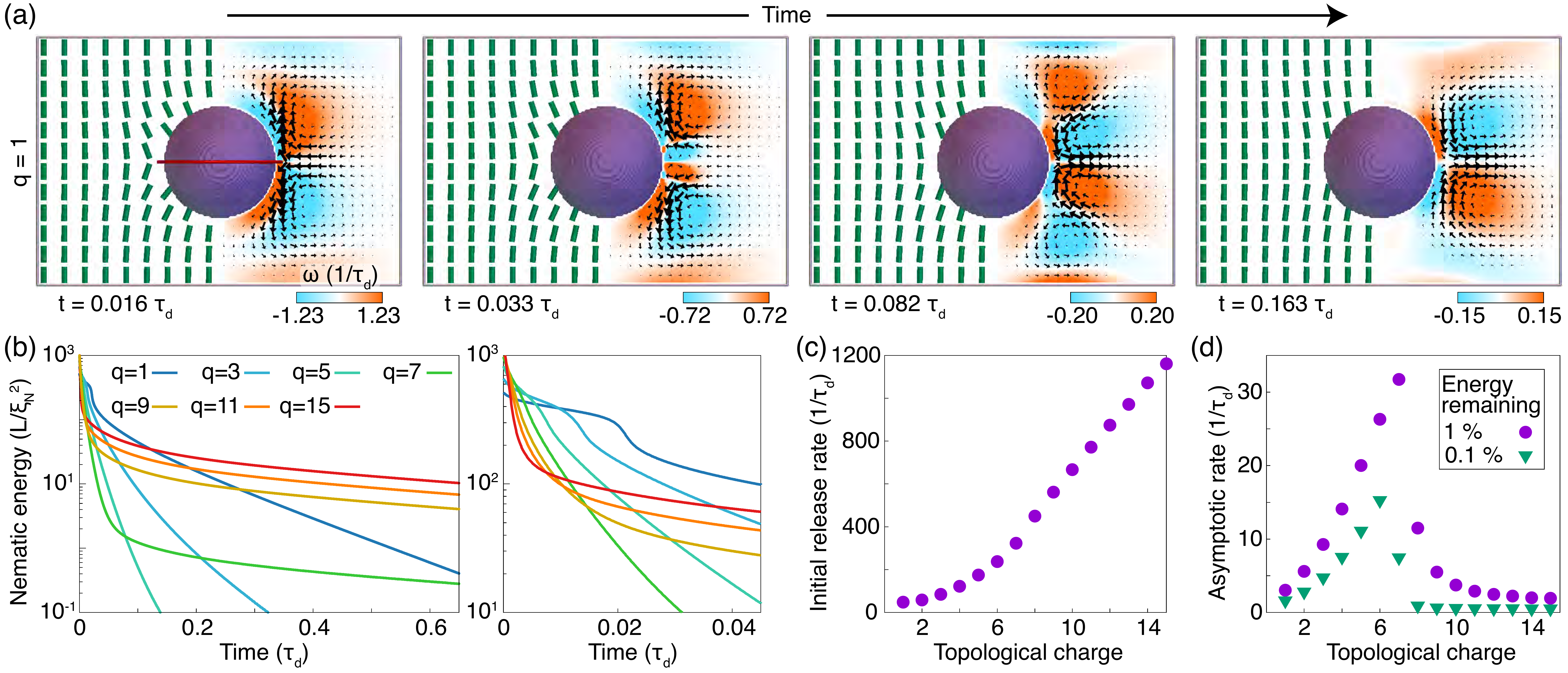}
	\caption{Discharging nematic topological batteries.
		(a)~As the orientational field (green rods) relaxes by changing its surface anchoring, the defect ring (red loop) is annhilated and drives a flow (black arrows) which changes direction over time. The flow vorticity $\omega$ measures the directon and magnitude of the flow.
		(b)~Higher $q$ states have both larger initial energy and faster initial discharge.
		(c, d)~The initial energy release rate, defined as the reciprocal of the $50\,\%$ energy discharge time, is monotonic with topological charge. In contrast, the asymptotic release rates (d), given by the reciprocals of the $99\,\%$ or $99.9\,\%$ discharge times, display topological resonance. See SI for simulation parameters.
	} 
	\label{nem2}
\end{figure*}

\subsection{Charging}

Nematic topological batteries are charged by imposing a topologically non-trivial director field anchoring profile on the surface of a spherical particle given by $\vec{n}_q(\theta,\phi)=\left(\sin\theta\cos q\phi, \sin\theta\sin q\phi, \cos\theta\right)$. Here, $\theta$ and $\phi$ are the polar and azimuthal angles, and the integer $q$ sets the effective topological charge of the particle~\cite{AlexanderGP_RevModPhys84_2012}. The system is placed in a cell with strong perpendicular anchoring of the nematic director on the top and bottom boundary (Fig~\ref{nem1}a). The bulk nematic structure counteracts the topological charge of the particle by forming a defect ring~\cite{cates2010}, known as a Saturn ring for $q=1$. The director field in the cross-sections of the defect ring varies continuously between $+1/2$, $-1/2$, and twisted profiles (Fig~\ref{nem1}a). For small $q$, the defect ring lies on the equatorial plane of the particle (Fig~\ref{nem1}a,b). As $q$ increases, the gradients of the director field increase, eventually causing the defect loop to buckle at a critical $q_c$ depending on colloid size. The resulting oscillatory defect profiles have wavelength matching that of the anchoring profile, and amplitude increasing with~$q$ (Fig~\ref{nem1}b). The buckling transition is characterized by the average horizontal director field, $\langle\mathbf{n}_{\parallel}\rangle$, which is zero for $q< q_c$ and positive for $q\geq q_c$ (Fig~\ref{nem1}c,d). The total free energy, $F$, however, strictly increases with~$q$ (Fig~\ref{nem1}e). Topology thus determines the distribution of energy throughout different modes, along with the total energy.

\subsection{Discharging dynamics and topological resonance}

Energy partitioning within the charged nematic batteries causes topological resonances in the discharge dynamics. The relaxation process is triggered by switching off the anchoring on the spherical particle, allowing the director field to take any orientation at its surface (Fig~\ref{nem2}a). The initial reorientation of the director field is accompanied by the shrinking of the defect ring until it is annihilated on the particle surface. The continued reorientation of the director field towards the equilibrium homogeneous vertical structure drives a complex flow pattern (Fig~\ref{nem2}a). During the typical director relaxation time, $\tau_\text{d}$, the energy decreases by several orders of magnitude (Fig~\ref{nem2}b). While the initial energy release rate increases with initial battery energy and topological charge~$q$ (Fig~\ref{nem2}c), the asymptotic release rates show characteristic resonance peaks (Fig~\ref{nem2}d). In particular, we observe superfast discharge rates for $q$ close to~$q_c$ (Fig~\ref{nem2}d). The return to slow relaxation for  $q\gg q_c$ is caused by the formation of the  long-scale deformation mode with $\langle \vec{n}_\parallel\rangle>0$ (Fig~\ref{nem1}c,d), which obstructs fast discharge.

\par
\section{Conclusions}

In both studied systems, topology and mechanics interact to produce long-lived states that are central to the observed topological resonances. For elastic batteries, these states are the intermediate knots that can form as the fiber unties. At small or zero twist, these configurations trap bending energy, leading to long relaxation times. Similarly, at sufficiently high topological charge, nematic batteries can store energy in the slow $\langle \vec{n}_\parallel\rangle$ mode. Over long timescales, these modes make the pivotal contribution to the overall discharge rate. More generally, this principle of topologically activated slow modes can provide a conceptual framework for understanding topological resonances in other soft matter settings.

\par 

To conclude, knotted elastic fibers and topologically charged nematic fluids provide prime demonstrations of topological batteries. Both systems permit controlled triggering of energy release via fracture and photoalignment techniques. Our above analysis shows how elastic and nematic batteries can be topologically optimized to achieve slow or fast discharge. Owing to the inherent robustness of topological structures, the above ideas translate to a wide range of scales. The energy stored in topological batteries may be harnessed to drive flows or perform mechanical or electrical work, thus opening an avenue for topological control of soft systems.
\par
This work was supported by a MathWorks fellowship (V.P.), the Slovenian Research Agency (ARRS) through grants  L1-8135 (M.R.), P1-0099  (M.R. and  \v{Z}.K.),  and N1-0124 (\v{Z}.K.), by the James S. McDonnell Foundation (J.D.) and the Robert E. Collins Distinguished Scholar Fund (J.D.).


%


\newpage

\newpage

\end{document}